\documentclass[aoas,preprint]{imsart}

\RequirePackage[OT1]{fontenc}
\RequirePackage{amsthm,amsmath,amsfonts}
\RequirePackage{natbib}
\RequirePackage[colorlinks,citecolor=blue,urlcolor=blue]{hyperref}
\usepackage{url}
\usepackage{graphicx}
\usepackage{subfig}
\usepackage{float}
\usepackage[normalem]{ulem}
\usepackage{enumitem}
\usepackage{array}
\arxiv{arXiv:0000.0000}
\usepackage{color}

\startlocaldefs
\numberwithin{equation}{section}
\theoremstyle{plain}

\endlocaldefs

\newcolumntype{M}{>{\centering\arraybackslash}m{\dimexpr.25\linewidth-2\tabcolsep}}

\newcommand\redout{\bgroup\markoverwith
{\textcolor{red}{\rule[0.5ex]{2pt}{0.8pt}}}\ULon}

\newcommand{\indep}{\rotatebox[origin=c]{90}{$\models$}}
\newcommand{\PM}{PM$_{2.5} $}

\definecolor{mypink1}{rgb}{0.858, 0.188, 0.478}

\begin{document}

\begin{frontmatter}
\title{Causal inference in the context of an error prone exposure: air pollution and mortality}
\runtitle{Causal inference in error prone exposure}

\begin{aug}
\author{\fnms{Xiao} \snm{Wu}\thanksref{m1}}\ead[label=e1]{wuxiao@g.harvard.cedu}
,
\author{\fnms{Danielle} \snm{Braun}\thanksref{m1}}
\ead[label=e2]
{dbraun@mail.harvard.edu}
,
\author{\fnms{Marianthi-Anna} \snm{Kioumourtzoglou}\thanksref{m2}}
\ead[label=e3]{mk3961@cumc.columbia.edu}
,
\author{\fnms{Christine} \snm{Choirat}\thanksref{m1}}
,
\author{\fnms{Qian} \snm{Di}\thanksref{m1}}
\and
\author{\fnms{Francesca} \snm{Dominici}\thanksref{m1}}

\runauthor{X. WU et al.}

\affiliation{Harvard T.H. Chan School of Public Health\thanksmark{m1} and Columbia University Mailman School of Public Health\thanksmark{m2}}

\address{Harvard T.H. Chan School of Public Health \\Department of Biostatistics  \\677 Huntington Avenue\\ Boston, MA 02115\\
\printead{e1} \\
\printead{e2}}
\address{Columbia University\\
Mailman School of Public Health\\
Department of Environmental Health Sciences\\
722 W 168th Street\\
New York, NY 10032 \\
\printead{e3}}
\end{aug}

\begin{abstract}
We propose a new approach for estimating causal effects when the exposure is measured with error and confounding adjustment is performed via a generalized propensity score (GPS). Using validation data, we propose a regression calibration (RC)-based adjustment for a continuous error-prone exposure combined with GPS to adjust for confounding (RC-GPS). The outcome analysis is conducted after transforming the corrected continuous exposure into a categorical exposure. We consider confounding adjustment in the context of GPS subclassification, inverse probability treatment weighting (IPTW) and  matching. In simulations with varying degrees of exposure error and confounding bias, RC-GPS eliminates bias from exposure error and confounding compared to standard approaches that rely on the error-prone exposure.
We applied RC-GPS to a rich data platform to estimate the causal effect of long-term exposure to fine particles (\PM) on mortality in New England for the period from 2000 to 2012. The main study consists of $2,202$ zip codes covered by $217,660$ $1\textmd{km} \times 1\textmd{km}$ grid cells with yearly mortality rates, yearly \PM\ averages estimated from a spatio-temporal model (error-prone exposure) and several potential confounders. The internal validation study includes a subset of 83 $1\textmd{km} \times 1\textmd{km}$ grid cells within 75 zip codes from the main study with error-free yearly \PM\ exposures obtained from monitor stations. Under assumptions of non-interference and weak unconfoundedness, using matching we found that exposure to moderate levels of \PM \ ($8 <$ \PM\ $\leq 10\ {\rm \mu g/m^3}$) causes a 2.8\%  (95\% CI: 0.6\%, 3.6\%) increase in all-cause mortality compared to low exposure (\PM\ $\leq 8\ {\rm \mu g/m^3}$).
\end{abstract}



\end{frontmatter}

\newpage
\section{Introduction}

When trying to estimate exposure effects, observational studies are widely used but are susceptible to some well-recognized sources of bias, including but not limited to 1) exposure measurement error and 2) confounding. The measurement error can arise from using mismeasured exposures in the analysis, since obtaining estimates of the error-free exposures is not always feasible. In addition, there is a confounding problem due to the lack of randomization in observational studies. 

Measurement error approaches have been extensively studied in regression problems
\citep{fuller2009measurement}.
There is a large literature on this topic both in linear and nonlinear regression models, including likelihood-based approaches, regression calibration, simulation extrapolation (SIMEX), and Bayesian approaches
\citep{carroll2006measurement}. In addition, in the context of air pollution, which is the motivation of this work (see Section~\ref{datapp}), methods to adjust for measurement error under a non-causal framework have been previously proposed;
\cite{dominici2000measurement,van2008traffic,gryparis2008measurement,szpiro2011efficient,hart15_error,alexeeff2016spatial}. 
Many of them consider a ``widely used, effective [and] reasonably well-investigated''
\citep{pierce2004adjusting}
method to adjust for measurement error, namely regression calibration. This method utilizes the following combined study designs: a large main study, for which $W$ (the error-prone exposure)  and $\mathbf{D}$ (a set of error-free covariates) are observed, and a smaller validation study, for which in addition to observing $(W, \mathbf{D})$, $X$ is also observed (the error-free exposure). The basic idea of regression calibration is to fit the regression model $X|W,\mathbf{D}$ in the validation study, and use the coefficients from this model to predict $X$ in the main study. After this prediction step, the proposed statistical analysis is performed on the main study with the predicted error-free exposures, $\hat{X}$, to obtain parameter estimates, and either bootstrap or the sandwich variance estimation are used to obtain adjusted standard errors.
The simplicity of this algorithm disguises its power. 

In addition, observational studies are susceptible to confounding bias by factors that are associated with both the exposure and outcome of interest. Failure to account for them in the analysis may lead to substantial bias. 
Although most studies adjust for confounding, many do so by simply including the potential confounders as covariates in the outcome model.
However, doing so may lead to model misspecification and allows for residual confounding \citep{rothman2008modern}.
Therefore, addressing confounding bias in a causal inference framework can be advantageous. A common approach for confounding adjustment in this framework is using propensity scores, the probability of a unit being assigned to a particular treatment, or exposure in our setting, given the pretreatment confounders.

Using propensity scores to adjust for confounding in a causal inference framework
was first introduced by \cite{rosenbaum1983central}. After this seminal paper, advanced propensity score techniques, both for estimation and implementation, have been developed to estimate causal effects in observational studies -- for propensity score estimation see \cite{dehejia1998propensity,dehejia1999causal,mccaffrey2004propensity}; for propensity score implementation see \cite{hirano2003efficient,robins2000marginal,rosenbaum1984reducing,harder2010propensity}. A common technique for estimation of the propensity score is by fitting a logistic regression model to predict the treatment (or in our setting exposure) with potential confounders included as predictors in the model. Three common techniques for propensity score implementation are matching, inverse probability of treatment weighting (IPTW), and subclassfication \citep{harder2010propensity}. This traditional propensity score framework is only able to handle binary exposures.

In some cases the interest is in estimating the exposure effect for a categorical exposure. To handle categorical exposures, a generalized propensity score (GPS) framework has been developed \citep{imbens2000role}. \cite{imbens2000role} developed a natural analogue to propensity score estimation under categorical exposures, which uses multinomial logistic regression, instead of logistic regression, to predict multiple exposure categories 
with all potential confounders included as predictors. They describe an analogue to IPTW for categorical exposures. Although there is no natural analogue to matching and subclassification for the GPS 
\citep{imbens2000role, lechner2001identification, rassen2013matching}, \cite{yang2016propensity} propose an alternative way to estimate causal effects using matching and subclassification for a categorical exposures by averaging potential outcomes separately for each of the exposure categories.

Measurement error adjustment for binary exposures under a causal inference framework has been studied by \cite{babanezhad2010comparison} and \cite{braun2017propensity}. In \cite{babanezhad2010comparison}, authors investigate how mismeasured exposures impact the estimation of the causal effects using four different approaches to adjust for confounding, including ordinary least squares (OLS), IPTW, G-estimation and propensity score covariate adjustment.
They derive the asymptotic bias for these four estimators, and show they are equally affected by measurement error under linear models when exposure measurement error is independent of the confounders, but not otherwise. \cite{braun2017propensity} proposes a two-step maximum likelihood approach using validation data to adjust for the measurement error, which effectively corrects for measurement error in binary exposures under a causal inference framework. Specifically, they first use a likelihood based adjustment to correct for measurement error in the propensity score model and estimate an adjusted propensity score. Next, based on the adjusted propensity score, they perform a likelihood-based adjustment on the outcome model to adjust for measurement error in the exposure variable directly. These approaches, however, assume binary exposures and are not directly applicable to a categorical exposure.

In this work, we focus on settings for which we have a continuous exposure measured with error, yet our interest is in estimating causal effects on a categorical scale. We propose a regression calibration (RC)-based adjustment to adjust for the measurement error in the exposure combined with GPS to adjust for confounding (RC-GPS). The RC model is fitted using the continuous
exposure, regressing the true exposure on the error-prone exposure and additional covariates on which the measurement error could depend. Outcome analysis adjusting for confounding using GPS is then conducted after transforming the corrected continuous exposure into a categorical exposure. The proposed method is innovative in the following ways: 1) it provides a correction for measurement error in the exposure for both design and analysis phases with GPS; 2) GPS implementations can be paired with any GLM outcome model (e.g. log-linear model); 
3) we show how standardized bias can be used to assess balance in the context of GPS analysis for categorical exposures. 


In Section~\ref{methods} we introduce the proposed adjustment. We then run extensive simulations to assess the performance of our proposed adjustment in Section~\ref{simulations}. We apply our proposed approach to investigate the effect between long-term \PM\ exposure and mortality in New England (VT, NH, CT, MA, RI and ME), using zip code aggregated data from Medicare. For the entire Medicare population (main study), long-term exposure to fine particles (\PM) is determined from a spatio-temporal model that uses multiple different sources as input (meteorological, land use variables, satellite data, etc.). \PM\ exposure based on these predictions is inaccurate, but for a subset of zip codes (validation study) we have actual \PM\ concentrations measured at monitors (error-free exposure). Although in reality \PM\ concentrations measured at monitors could still contain measurement error, e.g. instrumental measurement error, for the purpose of this manuscript we use the word ``error-free'' to refer to \PM\ concentrations measured at monitors, since these are the best available source for ground-level \PM\ concentrations. 
Using this internal validation study, we apply our proposed RC-GPS in Section~\ref{datapp}. Finally, we conclude with a discussion in Section~\ref{disc}.

\section{Methods}  \label{methods}

\subsection{General Notation and Overview}
Let $Y$ denote the observed outcome, $X$ denote the corresponding true continuous exposure, $X_{c}$ denote the true categorical exposure, which is obtained from $X$ based on pre-specified cut-offs, 
selected according to scientific interest,
 $W$ denote the error-prone continuous exposure, $W_{c}$ denote the error-prone categorical exposure, which is obtained using the same pre-specified cut-offs, $\mathbf{D}$ denote error-free covariates associated with the measurement error, and $\mathbf{C}$ denote error-free confounders associated with the true exposure and outcome. 
There is no restriction on whether $\mathbf{D}$ and $\mathbf{C}$ include the same covariates or not. 
For the main study, only $(Y, W, \mathbf{D}, \mathbf{C})$ are observed. In addition, suppose a validation study for which  $(X, W, \mathbf{D})$ are observed. Note the validation study does not have to be internal. 

Our interest is in estimating the causal effect of a categorical exposure on the outcome in observational studies. The target estimand is the average treatment effect (ATE). 
Following the potential outcomes framework \citep{rubin1974estimating}, we assume no-interference \citep{cox1958planning}, which is sometimes referred to as the stable-unit-treatment-value assumption (SUTVA) \citep{rubin1990formal}. Under this assumption, we assume that the potential outcome for a given observation is not affected by the exposure of any other unit, and that each exposure defines a unique outcome for each observation. 

Furthermore, under overlap and weak unconfoundedness assumptions (discussed in detail later), GPS can be used to estimate the ATE with observed categorical exposures adjusting for confounding. In the main study, the exposure is mismeasured; only $W$ (along with $W_{c}$) are observed instead of $X_{c}$. Estimating the ATE based on $W_{c}$ instead of $X_{c}$ may result in biased estimates of the ATE. 
Our goal is to adjust for the measurement error in the exposure and obtain unbiased estimates of the ATE. We accomplish this by introducing a regression calibration (RC)-based adjustment for mismeasured exposures combined with GPS to adjust for confounding (RC-GPS). This approach relies on a main study/validation study design.
\subsection{Regression Calibration\label{rc}}
In this section, we propose a regression calibration approach to adjust for measurement error in a continuous exposure. 
The adjustment relies on two common measurement error assumptions: 1) Transportability: we assume that the relationship between $X$ and $W,D$ 
would be the same in the validation study where $X$ is observed and in the main study in which it is not. 2) Non-differential measurement error: $Y \indep W | X, \mathbf{D}$. This assumption is equivalent to the surrogacy assumption and it means the conditional distribution of outcome Y given $(W, X, \mathbf{D})$ depends only on $(X, \mathbf{D})$.  

The relationship between true exposures $X$ and error-prone exposures $W$, conditional on other covariates $\mathbf{D}$, is modeled using a regression model specified by mean and variance; 
\begin{equation} 
\begin{split}
E(X|W, \mathbf{D}) &= m_X(W, \mathbf{D},\boldsymbol{\gamma})  \\
Var(X|W, \mathbf{D}) &= V(W, \mathbf{D},\boldsymbol{\gamma}) \Sigma_{X|W, \mathbf{D}} V^T(W, \mathbf{D},\boldsymbol{\gamma}).
\end{split}
\end{equation}
Under transportability, we assume that the coefficients $\boldsymbol{\hat{\gamma}}$ which are estimated in the validation study are transportable to the main study. Thus, unobserved $X$ in the main study can be estimated using $m_X(W, \mathbf{D},\boldsymbol{\hat{\gamma}})$.
A well-studied case is a linear regression model specified by: 
\begin{equation} \label{eq1}
\begin{split}
E(X|W, \mathbf{D}) &= \gamma_0 + \gamma_1W + \boldsymbol{\gamma_{2}}^T \mathbf{D} \\
Var(X|W, \mathbf{D}) &= \Sigma_{X|W, \mathbf{D}}
\end{split}
\end{equation}

Under transportability, unobserved $X$ in the main study can be estimated using Equation~\ref{eq1};
\begin{equation*}
\hat{X} = \hat{\gamma_0} + \hat{\gamma_1}W + \boldsymbol{\hat{\gamma_{2}}}^T \mathbf{D}.
\end{equation*}
When $\text{tr}(\Sigma_{X|W, \mathbf{D}})=0$, this reduces to the standard regression calibration model in which we only need to estimate $E(X|W, \mathbf{D})$ and $\hat{X}$ is an unbiased estimator of $X$ \citep{carroll2006measurement}. \citet{carroll2006measurement} proved by Taylor series expansions that $\hat{X}$ is approximately unbiased when the model has good fit, i.e. $\text{tr}(\Sigma_{X|W, \mathbf{D}})$ is small. 



\subsection{Generalized Propensity Scores Estimation\label{est}}
In this section, we discuss the generalization of the propensity score, introduced in the causal literature by \cite{rosenbaum1983central} for a binary treatment, to the setting of categorical exposures. We follow the generalization proposed by \cite{imbens2000role}. Under the assumption that we know the true exposure, denote $X_c \in \mathbb{X}_c = \{1,2,...,n\}$ the true categorical exposure having $n$ categories.
Let  $p(x|{\bf c}) = Pr(X_c=x|{\bf C}={\bf c})$  
for pre-specified categories $x = 1,2,3,\dots,n$.
We define the GPS as the conditional probability of receiving each category of the exposure given other pre-exposed covariates ${\bf c}$:
\begin{equation}
GPS({\bf c}) = (p(1|{\bf c}), p(2|{\bf c}),..., p(n|{\bf c}))
\end{equation}
The individual $p(x|{\bf c})$ is called the $x$-th element of $GPS({\bf c})$.

To model $GPS({\bf c}) = (p(1|{\bf c}), p(2|{\bf c}),..., p(n|{\bf c}))$, we consider a generalized linear model (GLM) relating $X_{c}$ 
to $\bf C$, that is $p(x|{\bf c}) = Pr(X_{c}=x|{\bf C}={\bf c},\boldsymbol{\eta}) = g^{-1}(\eta_{0x} +\boldsymbol{\eta_{1x}}^T {\bf c})$, where $g$ is known. One common $g$ is the multinomial logistic regression model.
\begin{equation*}
\ln \frac{Pr(X_{c}=x|{\bf C}={\bf c},\boldsymbol{\eta})}{Pr(X_{c}=n|{\bf C}={\bf c},\boldsymbol{\eta})} = \eta_{0x} +\boldsymbol{\eta_{1x}}^T {\bf c}.
\end{equation*}

\subsection{Generalized Propensity Scores Implementation\label{imp}}
We consider three GPS implementations; subclassification, IPTW and matching, all conditional on the estimated GPS \citep{yang2016propensity,imbens2000role}. Following \cite{yang2016propensity}, let $X_{c,j}$ denote the true categorical exposure for unit $j$, $X_{c,j} \in \mathbb{X}_c = \{1,2,...,n\}$, and $Y_{j}(x)$ denote the potential outcome for an exposure $x$ for observation $j$. The observed outcome can then be written as $Y_j^{obs} = Y_{j}(X_{c,j})$. Define the indicator variables $I_j(x) \in \{0,1\}$,
\begin{equation*}
I_{j}(x) =
\left\{
\begin{aligned} 
&1 \ \ \text{if} \ X_{c,j}=x, \\
&0 \ \ \text{otherwise}.
\end{aligned}
\right.
\end{equation*}
In addition to the no-interference assumption described above, we require the following two assumptions for proposing the GPS implementations. 

\textbf{Assumption 1} (Overlap/Positivity) For all values of $\mathbf{c}$, the probability of receiving any category of the exposure is positive:
\begin{equation*}
Pr(X_c=x|\mathbf{C}=\mathbf{c}) > 0 \ \text{for all} \ x, \ \mathbf{c}
\end{equation*}
This assumption guarantees that for all possible values of $\mathbf{c}$, we will be able to estimate the ATE for each category of the exposure without relying on extrapolation. In many applications, there are regions of the confounder space with low probability values of receiving one of the exposures, which leads to a violation of this assumption.  There are methods for improving overlap; specifically, both \cite{harder2010propensity} and \cite{yang2016propensity}
suggest dropping units from the analysis with low and high values of the GPS, and conducting analysis on the trimmed sample.

\textbf{Assumption 2} (Weak Unconfoundedness) The assignment mechanism is weakly unconfounded if for all $x \in \mathbb{X}_c = \{1,2,...,n\}$,
\begin{equation*}
I_j(x) \ \indep \ Y_j(x) \ | \ \mathbf{C}_j.
\end{equation*}
There are two things to note about this assumption. First, it can be preserved if we condition on a specific scalar function of $\mathbf{C}_j$, i.e. $p(x|\mathbf{C}_j)$, as shown in Lemma 1 below. This is favorable since it allows for the reduction in the dimension of the conditioning covariates when estimating causal effects. Second, this assumption is sufficient for constructing a form of the ATE, which will be formalized in Lemma 2. 

\textbf{Lemma 1} (Weak Unconfoundedness given GPS) Suppose the assignment mechanism is weakly unconfounded. Then for all $x \in \mathbb{X}_c = \{1,2,...,n\}$,
\begin{align*}
I_j(x) \ \indep \ Y_j(x) \ | \ p(x|\mathbf{C}_j).
\end{align*}
Lemma 1 allows us to estimate the following ATE, described in Lemma 2. 

\textbf{Lemma 2} (Average Treatment Effects under Weak Unconfoundedness) Suppose the assignment mechanism is weakly unconfounded. Then for all $x,x' \in \mathbb{X}_c = \{1,2,...,n\}$,
\begin{equation*}
\begin{split}
ATE(x';x) &= E[Y_j(x')-Y_j(x)]  \\
&= E \big[ E(Y_j^{obs}| X_{c,j}=x', p(x'|\mathbf{C}_j)] \big] - E \big[ E(Y_j^{obs}| X_{c,j}=x, p(x|\mathbf{C}_j)] \big].
\end{split}
\end{equation*}
Lemma 2 allows us to loosen the constraint in comparison of exposure effects. Instead of conditioning on the full set of $n-1$ generalized propensity scores $(p(1|\mathbf{C}_j),...,p(n-1|\mathbf{C}_j))$, we can estimate the average effect $E[Y_j(x')-Y_j(x)]$ by constructing an overall average estimate for each exposure category $x$ separately.
For a single exposure category $x$, the corresponding subpopulations are defined by the value of a single score, $p(x|\mathbf{C}_j)$, leading to the equality;
\begin{equation*}
 E[Y_j(x)] = E\big[E[Y_j^{obs}|X_{c,j}=x,p(x|\mathbf{C}_j)]\big].
\end{equation*}
Even though the comparisons of exposure effects are not constructed by conditioning on the full set of generalized propensity scores, which makes us lose the ability to create subpopulations where we can extrapolate causal effects, our estimated ATE for whole population under the weak unconfoundedness assumption is still valid in causal inference \citep{imbens2000role}. 
In contrast, under strong uncoundedness where $X_{c,j} \ \indep \ (Y_j(1),Y_j(2),...,Y_j(n)) \ | \ \mathbf{C}_j$  (defined in \cite{rosenbaum1983central}), we can estimate ATE for subpopulation, i.e. the ATE in population with exposure category 1 and 2 only. 
Our interest, however, is usually in causal effects for whole population, which can be achieved under the weak unconfoundedness assumption.


\subsubsection{Subclassification} We follow the approach proposed by \cite{yang2016propensity}. Consider classifying individuals into $K$ groups based on the $x$-th GPS element, each group containing $N_{k,x}$ observations having similar values of the corresponding estimated GPS elements. The most common way to construct subclasses is to use quantiles of the GPS. The ATE between two exposures, $x'$ and $x$, i.e. $ATE(x';x)$, can be written as the difference of two expectations $E[Y_j(x')]$ and $E[Y_j(x)]$, which can be estimated separately. Let $q_{x,k}^{p(x|\mathbf{c}_j)}$ be the value of $p(x|\mathbf{C}_j)$ in $k$-th quantile in the sample. The average value of $Y_j(x)$ in subclass $k$ is estimated as;
\begin{equation*}
\hat{\mu}_{k,x} = \frac{1}{N_{k,x}} \sum_{j: q_{x,k-1}^{p(x|\mathbf{c})} \leq p(x|\mathbf{C}_j) < q_{x,k}^{p(x|\mathbf{c})},X_j=x} Y_j^{obs}
\end{equation*}
where $N_{k,x}$ is the number of units 
with the $x$-th GPS element falling into the interval $[q_{x,k-1}^{p(x|\mathbf{c})}, \ q_{x,k}^{p(x|\mathbf{c})})$ and $X_{c,j}=x$. The overall average of $Y(x)$, $E[Y_j(x)]$, 
is then estimated as,
\begin{align*}
\hat{E}[Y_j(x)] &=\hat{E}\big[E[Y_j^{obs}|X_{c,j}=x,p(x|\mathbf{C}_j)]\big] \\
&= \sum^{K}_{k=1} \frac{N_k}{N} \hat{\mu}_{k,x}
 \end{align*}
where $N_{k}$ is the number of individuals with the $x$-th GPS element falling into the interval $[q_{x,k-1}^{p(x|\mathbf{c})}, \ q_{x,k}^{p(x|\mathbf{c})})$, and $N$ is the total sample size. We can estimate $E[Y_j(x')]$ similarly, and consequently, obtain the ATE between any two categories of exposure. 


\subsubsection{Inverse Probability of Treatment Weighting (IPTW)} IPTW involves weighting each individual by the inverse of their GPS. This approach was first introduced by \citet{imbens2000role} and is an analog of using IPTW with propensity scores under a binary exposure. The probability weight assigned to a particular individual is the GPS element corresponding to its true category of treatment. Note that the use of IPTW can be construed as a further extension of subclassification, with the number of subclasses going to infinity.  
One can estimate the overall average of $Y_j(x)$ using IPTW as;
\begin{align*}
\hat{E}[Y_j(x)] &= \hat{E}[ \frac{Y_j^{obs} I_j(x)}{p(x|\mathbf{C}_j)}].
\end{align*}
We can estimate $E[Y_j(x')]$ similarly, and consequently, obtain the ATE between any two categories of exposure.

\subsubsection{Matching} We follow the approach proposed by \cite{yang2016propensity}, which involves matching individuals who receive one category of exposure to individuals who received another category of exposure based on their estimated GPS. There are various ways of matching, e.g. matching on the full set of GPS, yet here we match based on a scalar variable to furthest reduce the dimensionality of the matching problem. 
Following \cite{yang2016propensity}, we define a one-to-one nearest neighbor matching function with replacement,
\begin{equation*}
	m_{gps}(x,p)=\text{arg} \ \text{min}_{j: X_{c,j} = x} ||p(x|\mathbf{C}_j) - p||.
\end{equation*}
Using this matching function, we impute $Y_j(x)$ as: $\hat{Y}_j(x)=Y_{m_{gps}(x,p(x|\mathbf{C}_j))}^{obs}$ for $j=1,2,...,N$ successively, to create a dataset with the sample size N, yet with replicated observations. By resampling with replacement from the original dataset, we can construct a finite sample representing the pseudo subpopulation having exposure $X_{c,j} = x$. The overall average of $Y_j(x)$ can be expressed as;
\begin{align*}
	\hat{E}[Y_j(x)] =\frac{1}{N}\sum_{i=1}^{N} Y^{obs}_{m_{gps(x, p(x|\mathbf{C}_i))}}.
\end{align*}
We can estimate $E[Y_j(x')]$ similarly by creating another dataset with observations receiving exposure $X_{c,j} = x'$ using the matching function defined above, and consequently, obtain the ATE between any two categories of exposure.

\subsection{Outcome Analyses}
In the causal framework, one might be interested in a specific statistical quantity, e.g.  ratio measures. 
The three GPS implementations are not explicit about the forms of the outcome model, and provide the flexibility to estimate such statistical quantities directly from the estimates of the overall averages for each exposure category. For example, the ATE measured by ratio can be expressed as;
\begin{align*}
ATE_{ratio}(x';x) &= \frac{\hat{E}[Y_j(x')]}{\hat{E}[Y_j(x)]}.
\end{align*}
However, if one is interested in incorporating covariates into the outcome model, to further adjust for confounding, one may want to specify an outcome model. For instance, one may specify the following outcome model; $Y|X_c,\mathbf{C}$ using a GLM, $E(Y(X_{c})) = r^{-1}(\beta_0+ \sum^{n}_{x=1} \beta_{1x} I(X_c =x)+\boldsymbol{\beta_2} \mathbf{C})$, where $I(\cdot)$ is an indicator for the corresponding exposure category. 
While we conduct all analyses assuming $\beta_2 = 0$ and
exclude $\mathbf{C}$ from the outcome model, $\beta_2\neq 0$ could be included to adjust for residual imbalances
not captured by the generalized propensity score implementation or to improve precision of causal estimates \citep{harder2010propensity}.

For IPTW, the outcome model can be easily implemented as a GLM with the corresponding GPS elements specified as weights. For subclassification, the outcome model is essentially implemented on samples selected from subclasses constructed by the corresponding GPS elements, 
and then weighted by sample size of their corresponding subclass. 
For matching, the outcome model is essentially implemented on the replicated samples constructed by GPS matching as described above.

For inference, we estimate the standard errors (SEs) of the ATE using bootstrap to jointly account for the variability in the estimation of RC parameters $\boldsymbol{\gamma}$, GPS parameters $\boldsymbol{\eta}$, and outcome model parameters $\boldsymbol{\beta}$. We use standard bootstrap to construct the SEs for GPS subclassfication and IPTW. We use a modified bootstrap method for GPS matching, as standard bootstrap may provide invalid standard errors for matching \citep{abadie2008failure}. Certain modifications to the standard bootstrap, like the m-out-of-n bootstrap in \cite{bickel2012resampling}, were proven to recover the validity. 



\subsection{Proposed RC-GPS}
\label{outline}
Our proposed RC-GPS adjustment is a two-stage approach. \\ 
\textbf{Stage 1: Measurement Error Correction}
\begin{enumerate}[leftmargin=0.4cm]
  \item Fit a RC model in the validation study. More specifically, fit $E(X|W,\mathbf{D}) = \gamma_0 +\gamma_1 W + \boldsymbol{\gamma_2}^T \mathbf{D}$ to obtain estimated $\boldsymbol{\gamma}$, i.e. $\boldsymbol{\hat{\gamma}}$ in the validation study. The form of RC model is not restricted to linear regression, although otherwise one needs further justifications of approximations for the RC model to fully adjust for measurement error (Section~\ref{rc}). 
  \item Under the transportability assumption, estimate $\hat{X} = \hat{\gamma_0} + \hat{\gamma_1}W + \boldsymbol{\hat{\gamma_{2}}}^T \mathbf{D}$ in the main study. The $\hat{X}$ is approximately unbiased if the RC model is correctly specified and has good fit (i.e $\text{tr}(\Sigma^2_{X|W,\mathbf{D}})$ is small). 
  \item Based on pre-defined categories, transform $\hat{X}$ into $\hat{X}_{c} \in \mathbb{X}_c = \{1,2,...,n\}$, a categorical variable. The choice of category can be determined to either be policy-relevant or by optimizing overlap through sensitivity analyses. 
\end{enumerate}
\textbf{Stage 2: GPS Estimation, Implementation, Outcome Analysis}\\
\textit{Stage 2A: Design Phase with GPS} 
\begin{enumerate}[leftmargin=0.4cm]
\setcounter{enumi}{3}
\item After obtaining $\hat{X}_{c}$ in the main study, 
estimate the GPS model using a GLM relating $\hat{X}_{c}$ to $\mathbf{C}$ as described in Section~\ref{est}. The estimated GPS is approximately error-free if the RC model is correctly specified and has good fit.
\end{enumerate}
\textit{Stage 2B: Analysis Phase with GPS}
\begin{enumerate}[leftmargin=0.4cm]
\setcounter{enumi}{4}
\item Estimate $\hat{E}[Y(x)]$ for each exposure category $x \in \mathbb{X}_c = \{1,2,...,n\}$
after adjusting for confounding using GPS subclassfication, IPTW or matching methods (Section~\ref{imp}).
\item Estimate the ATE as the contrast of $\hat{E}[Y(x)]$ and $\hat{E}[Y(x')]$ between any two exposure categories $x,x'$.   
\item Estimate the SEs of the ATE using bootstrap to jointly account for the variability in the estimation of RC parameters $\boldsymbol{\gamma}$, GPS parameters $\boldsymbol{\eta}$, and outcome model parameters $\boldsymbol{\beta}$.
\end{enumerate}

\section{Simulations\label{simulations}}
\label{sec3}


We conduct simulation studies to evaluate the performance of the proposed RC-GPS approach under the three types of GPS implementations outlined in Section~\ref{outline} (IPTW, subclassification, and matching). We estimate the ATE based on 1) the true exposures in both GPS and outcome models, 2) the error-prone exposures in both GPS and outcome models, 3) our proposed RC-GPS adjustment. 

\subsection{Simulation Strategies}
We generate a main study/internal validation study setting, in which the validation study is randomly sampled from the main study. 
The data generation strategy is summarized in Table~\ref{tablep}. Briefly, we generate six confounders $(C_1,C_2,...,C_6)$, which include a combination of continuous and categorical variables.  We generate three covariates in the measurement error model, $(D_1,D_2,D_3)$, which are continuous. Note that $\mathbf{C}$ and $\mathbf{D}$ could include the same covariates, although this is not required. For simulations we assume that $C_1=D_1$, but that the remaining covariates are different. 

The variables $[W|\mathbf{C},\boldsymbol{\tau}], [X|W,\mathbf{D},\boldsymbol{\gamma}], [Y|X,\mathbf{C},\boldsymbol{\beta}]$ were generated as continuous under linear regression models with parameters specified in Table~\ref{tablep}. 
We begin by generating the error-prone exposures $W$, and then generate $X$, which guarantees the correct specification of the RC model. 
We consider 7 settings, where we vary 1) $\boldsymbol{\tau}$ to control the strength of confounding for exposure, 2) $\gamma_1$ to control the correlations between $X$ and $W$, 3) $\Sigma_{X|W,\mathbf{D}}$ to control the goodness of  RC model fit, 4) quadratic term in the RC model to control RC model misspecification, 5) $\beta_1$ to control the magnitude of treatment effect in outcome model, and 6) $\boldsymbol{\beta}_{2}$ to control the strength of confounding for outcome. The default setting, discussed in detail in the main text, is highlighted in Table~\ref{tablep}. 
We fix the sample size of the main study as 2000 and the internal validation study as 500. We conduct 1000 replicates of each scenario. The R code for all simulations is available on github \href{https://github.com/wxwx1993/RC-GPS}%
{https://github.com/wxwx1993/RC-GPS}.
\begin{table}[h!]
\centering
\caption{Simulation parameters: data generating mechanism under which simulations were conducted.} 
\begin{tabular}{l l}
\hline
Confounding for exposure, $E[W| \mathbf{C}, \boldsymbol{\tau}]$& $\boldsymbol{\tau}$ \\  [1ex]
   (1) \textbf{Moderate confounding}  &
 $(0.8, 0.8, 1.6, 1.2, 2.4, 1.6, 2.4)^{T}$ \\ [1ex]
 (2) Large confounding  &
 $(1.6, 1.6, 3.2, 2.4, 4.8, 3.2, 4.8)^{T}$ \\ [1ex]
   \hline 
 Measurement Error Model, $E[X|W, \mathbf{D}, \boldsymbol{\gamma}]$& $\boldsymbol{\gamma}$ \\ [1ex] 
   (1) \textbf{Strong correlation: 0.85}  &$ \gamma_1=0.8,\boldsymbol{\gamma}_{2}=(2,1,3)^{T} $\\ [1ex]
   (2) Weak correlation: 0.40  &$ \gamma_1=0.2,\boldsymbol{\gamma}_{2}=(2,1,3)^{T} $\\ [1ex]
   \hline 
  Measurement Error Model Fit& $\Sigma_{X|W,\mathbf{D}}$ \\ [1ex] 
   (1) \textbf{Good of fit}  & diag$(\Sigma_{X|W,\mathbf{D}})= \boldsymbol{I}_N $\\ [1ex]
   (2) Lack of fit  & diag$(\Sigma_{X|W,\mathbf{D}})= 10 \boldsymbol{I}_N $\\ [1ex]
   \hline 
  Measurement Error Model Specification&  Model structure \\ [1ex] 
   (1) \textbf{Linear Model}  &$ X= \gamma_1 W + \boldsymbol{\gamma}_{2} \mathbf{D} $\\ [1ex]
   (2) Quadratic Model $(\gamma_3=0.05)$  &$ X= \gamma_1 W + \boldsymbol{\gamma}_{2}\mathbf{D} + \gamma_3 W^2 $\\ [1ex]
   \hline 
Outcome Model, $E[Y|X, \mathbf{C}, \boldsymbol{\beta}]$ & $\boldsymbol{\beta}$ \\ [1ex]
(1) \textbf{Large treatment effect}  & {$\beta_1=1,\boldsymbol{\beta}_{2}=(3,2,1,4,2,1)^{T}$} \\    [1ex]
(2) Small treatment effect  & {$\beta_1=0.5,\boldsymbol{\beta}_{2}=(3,2,1,4,2,1)^{T}$} \\    [1ex]
    \hline
Confounding for outcome & $\boldsymbol{\beta}$ \\ [1ex]
(1) \textbf{Moderate confounding}  & {$\beta_1=1,\boldsymbol{\beta}_{2}=(3,2,1,4,2,1)^{T}$} \\    [1ex]
(2) Large confounding  & {$\beta_1=1,\boldsymbol{\beta}_{2}=(15,10,5,20,10,5)^{T}$} \\    [1ex]
    \hline
Sample size & $N_{m}$ and $N_{v}$ \\  [1ex]
    (1) & $N_{m}= 2,000$ and $N_{v} = 500$\\   [1ex]
    \hline
    Covariate  & Distribution \\  [1ex]
    (1) $C_{1}-C_3/C_{4}/C_{5}/C_6$ & {${\footnotesize N\left[\left(\begin{array}{c}
0\\
0\\
0
\end{array}\right),\left(\begin{array}{ccc}
2 & 1 & -1\\
1 & 1 & -0.5\\
-1 & -0.5 & 1
\end{array}\right)\right]}/U\{-2,2\}/U(-3,3)/\chi^2(1)$} \\
    (2) $D_{1}/D_{2}/X_{3}$ & {$C_1/N(0,4)/U(-5,5)$} \\ [1ex]
    \hline
Cut-off points & ${\bf k}$ \\  [1ex]
    (1) & $k_1=-5$ and $k_2=15$\\   [1ex]
    \hline
\end{tabular}
\label{tablep}
\end{table}

\subsection{Simulation Results}
To implement the RC-GPS we follow the approach described in Section~\ref{outline}. After fitting the RC model and estimating $\hat{X}$ in the main study, we categorize these estimates into three categories based on pre-defined cut-off points $(k_1=-5, k_2=15)$, and obtain exposure categories $\hat{X}_c=1, 2, 3$. 
We also obtain $W_c$ from $W$ using the same cutoffs. Using $\hat{X}_c$ we then fit the GPS model and estimate the ATE. For subclassification, we classify subjects into ten subclasses by deciles based on each GPS element. For IPTW, weights were calculated as the inverse of the corresponding GPS elements as described in Section~\ref{imp}, and extreme weights are set equal to 10 if the weights are greater than 10 \citep{harder2010propensity}. 
For matching, we use a form of one-to-one nearest neighbor matching with replacement based on the corresponding GPS elements as described in Section~\ref{imp}. 

We provide a detailed description of the simulation results from the default setting (highlighted in bold in Table~\ref{tablep}).  Under this setting, the true ATE is $\boldsymbol{\beta_{1}}=(22.56,\ 21.50)$ (for exposure categories $\hat{X}_c=2$ vs. $1$ and $\hat{X}_c=3$ vs. $2$, respectively) which is estimated by fitting the linear model $Y=\beta_0+   \sum^{3}_{i=2} \beta_{1i} I(X_c =i) +\boldsymbol{\beta_2} \mathbf{C}$ for a large simulated dataset with sample size $N=10^6$.

\begin{figure}
\includegraphics[width=0.49\textwidth]{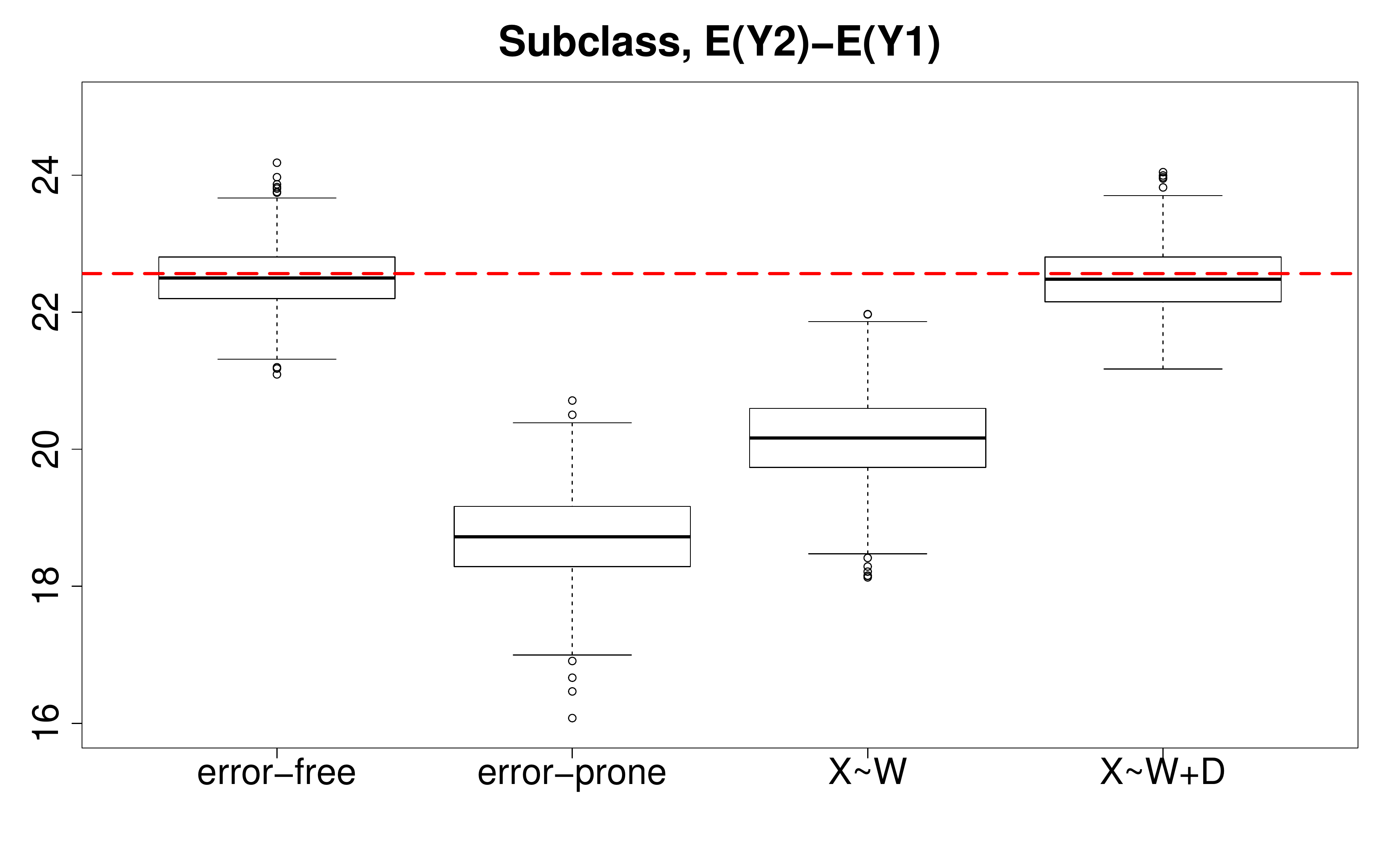}
\includegraphics[width=0.49\textwidth]{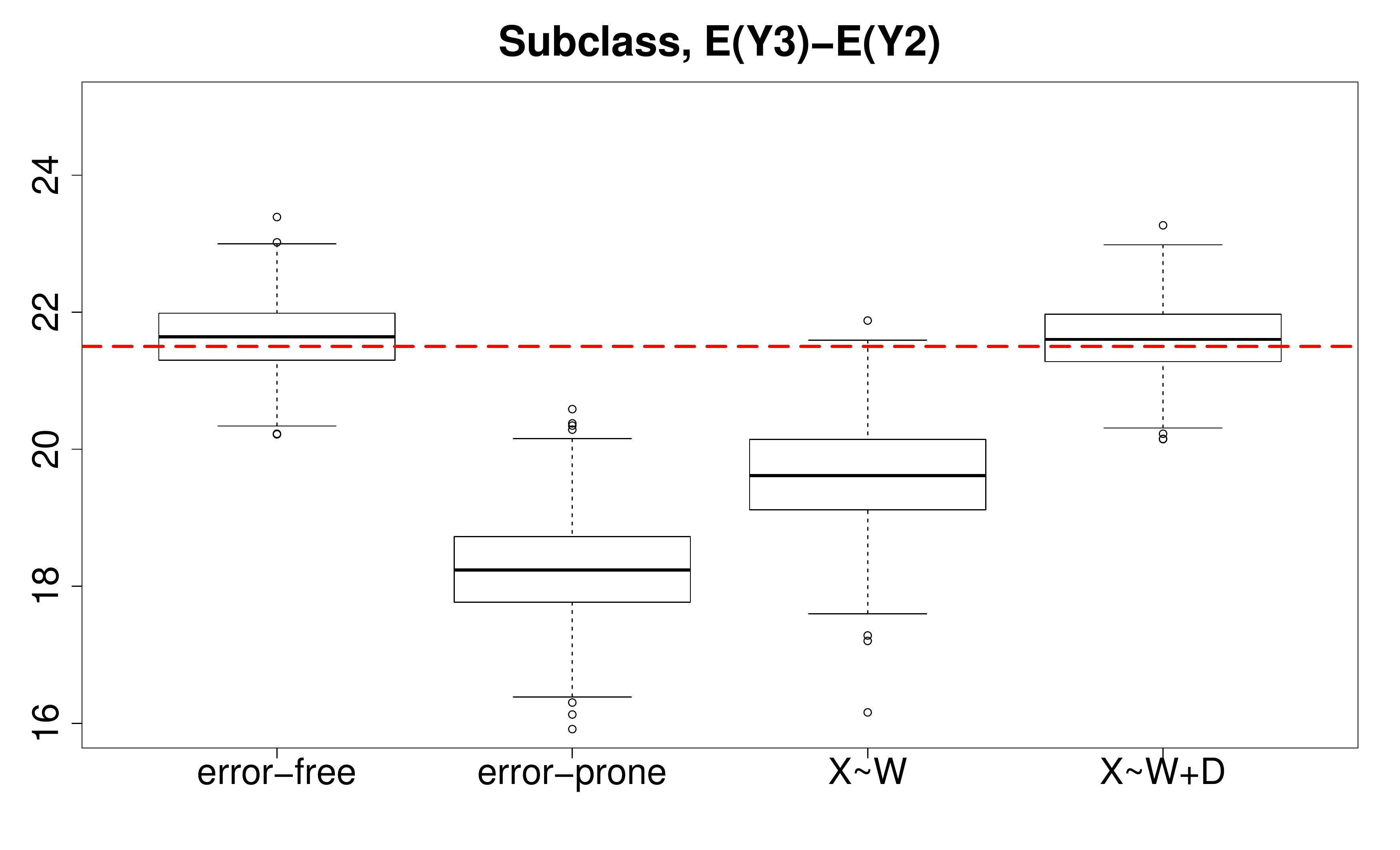}
\caption{Simulation results for default setting described in Table~\ref{tablep}. Subclassification: a) Error-free: GPS approach using error-free exposures $X_c$; b) Error-prone: GPS approach using error-prone exposure $W_c$; c) $X\sim W$: the proposed two-stage RC-GPS approach, in which exposures are estimated using a misspecified RC model $\hat{X} = \hat{\gamma_0} + \hat{\gamma_1}W$ which does not include covariates; d) $X\sim W+D$: the proposed two-stage RC-GPS approach, in which exposures are estimated using the correctly specified RC model $\hat{X} = \hat{\gamma_0} + \hat{\gamma_1}W + \boldsymbol{\hat{\gamma_{2}}}^T \mathbf{D}$ which includes covariates. The red dashed line represents the true ATE (gold standard).}
\label{figure2}
\end{figure}

The ATE for subclassification is shown in Figure~\ref{figure2}. 
The left plot represents the ATE of exposure $\hat{X}_c=2$ vs. $\hat{X}_c=1$, and the right plot represents the ATE of exposure $\hat{X}_c=3$ vs. $\hat{X}_c=2$. The ATE is estimated based on four different approaches from left to right: a) based on GPS approach using error-free exposure $X_c$ categorized from $X$; b) based on GPS approach using error-prone exposure $W_c$ categorized from $W$; c) based on the proposed two-stage RC-GPS approach, in which $\hat{X}$ is estimated using a misspecified RC model $\hat{X} = \hat{\gamma_0} + \hat{\gamma_1}W$ which does not include covariates; d) based on proposed two-stage RC-GPS approach, in which $\hat{X}_c$ is estimated using the correctly specified RC model $\hat{X} = \hat{\gamma_0} + \hat{\gamma_1}W + \mathbf{\hat{\gamma_{2}}}^T \mathbf{D}$ which includes covariates. The true ATE is denoted by the red dashed line. 

GPS implementation with subclassification using the error-free exposures results in a very small bias compared to the true ATE. Yet even in this setting where the error-prone and error-free exposures are highly correlated, GPS implementations using error-prone exposures result in significant bias, which illustrates the necessity of adjusting for the measurement error. RC-GPS using a correctly specified RC model performs really well, significantly reducing the bias of the estimated ATE compared to using the error-prone exposure. The bias was reduced from -17.07\% to -0.36\% and from -15.13\% to 0.55\% (exposure categories $\hat{X}_c=2$ vs. $1$ and $\hat{X}_c=3$ vs. $2$, respectively).
RC-GPS using a misspecified RC model still reduces, although does not completely eliminate, the bias. The bias under this setting was reduced from -17.07\% to -10.58\% and from -15.13\% to -8.77\% (exposure categories $\hat{X}_c=2$ vs. $1$ and $\hat{X}_c=3$ vs. $2$, respectively).
Additional results (Figure A1 in \ref{suppA}), show that IPTW and matching perform similarly compared to subclassification.

\subsection{Overlap and Balance} 
\label{simulations3}
\begin{figure}
\centering
\includegraphics[height=0.3\textheight,width=1\textwidth]{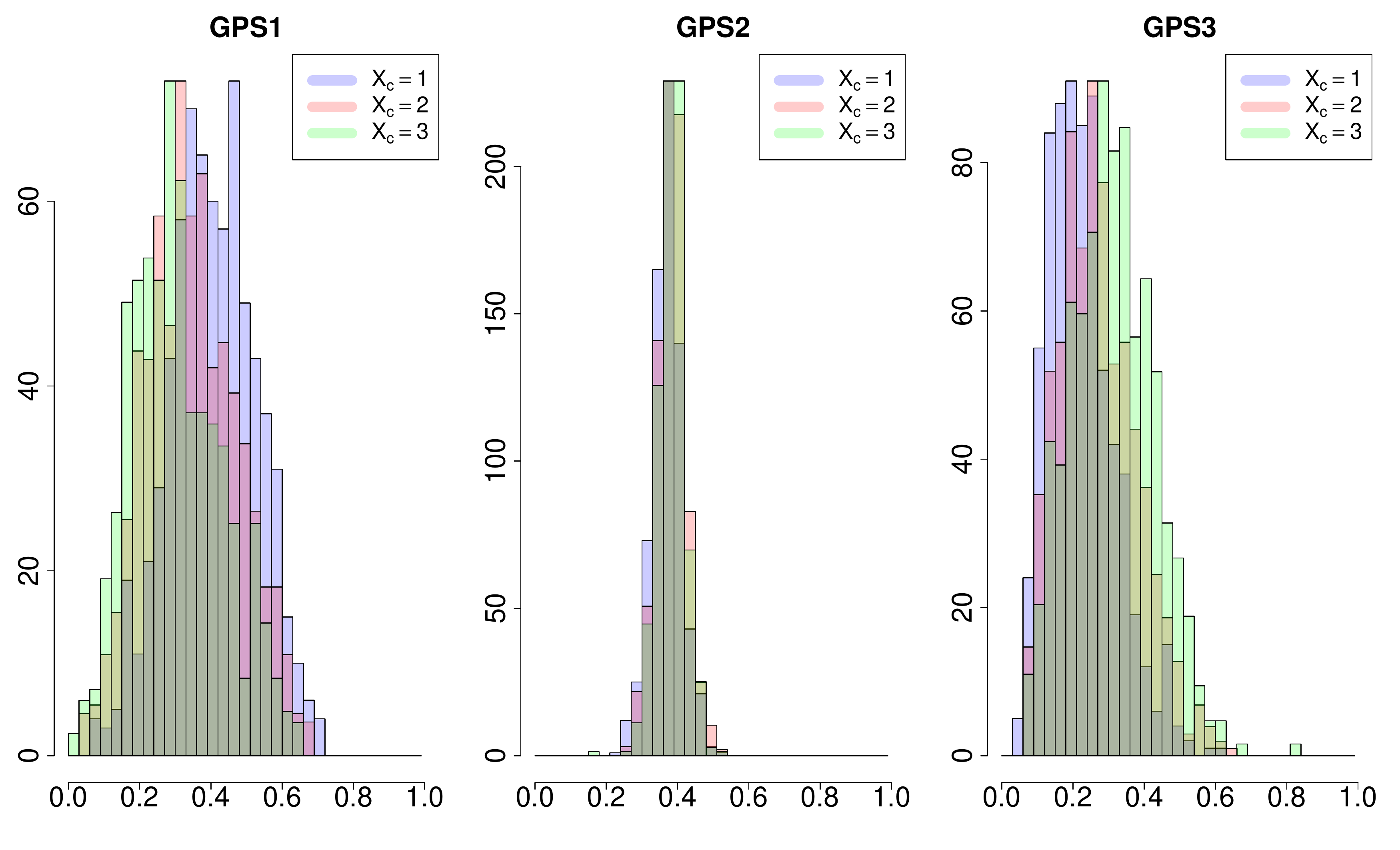}
\caption{Overlap. Each panel represents histograms of each corrected GPS element, colored according to different subpopulations. For assessing overlap,  we can see the majority of samples have GPS elements away from zero or one, and the ranges of each GPS element overlap across samples with different exposure providing the evidence that the overlap assumption likely holds. Results correspond to the default setting are described in Table~\ref{tablep}.}
\label{figure5}
\end{figure}

We evaluate the overlap assumption by inspecting the distributions of each estimated GPS element for all subjects as shown in Figure~\ref{figure5}. 
In this simulation study, we see that the overlap assumption likely holds in general, since the majority of samples have GPS elements away from zero or one. The histogram also shows the ranges of each GPS element overlap across samples with different exposure, which is referred to complete overlap in practice \citep{vaughn2008data}.

\begin{figure}[h]
\centering
\includegraphics[height=0.3\textheight,width=1\textwidth]{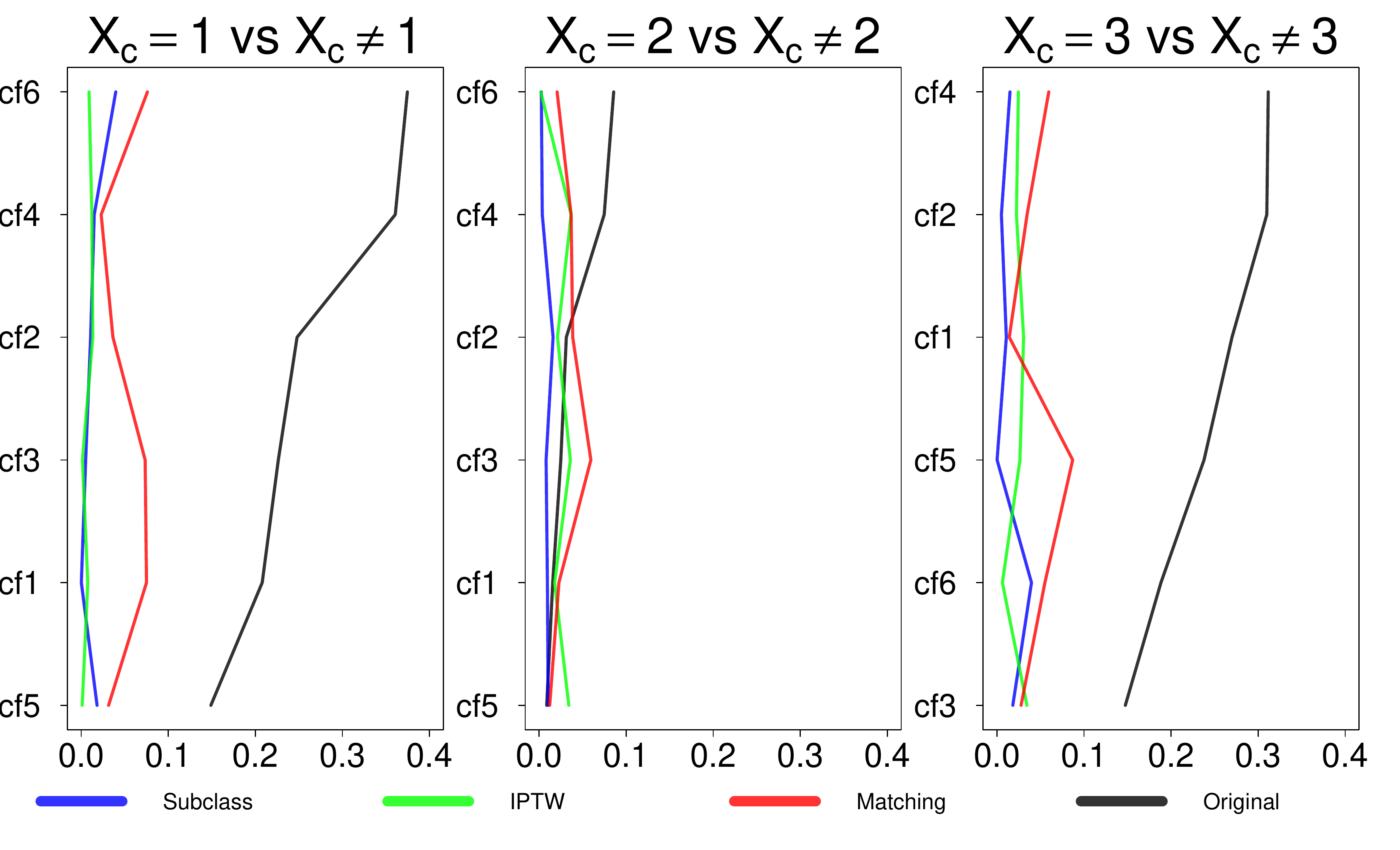}
\caption{Absolute Standardized Bias (ASB). Each panel represents the absolute standardized biases for each of the six confounders (cf), between subpopulations with $X_c=x$ and subpopulations with $X_c \neq x$ in the original data (black) and after GPS implementations (colored). All three GPS implementations perform similarly and all improve confounder balance substantially. Results correspond to the default setting are described in Table~\ref{tablep}.}
\label{figure4}
\end{figure}

Under weak unconfoundness, the single GPS element $p(x|\mathbf{c_j})$ is only required to achieve balance between subpopulations with $X_c=x$ and subpopulations with $X_c \neq x$. One can assess balance by estimating the absolute standardized biases of each confounder before and after GPS implementation using techniques similar to those described in \cite{harder2010propensity}. The key is that each GPS element $p(x|\mathbf{c_j})$ is treated as a binary exposure propensity score, and  balance is evaluated across all confounders between subpopulations with $X_c=x$ and subpopulations with $X_c \neq x$. For each of the six confounders, we estimate the absolute standardized bias (ASB). The ASB for each covariate is calculated by dividing the difference in means of the covariate between the treated group and the comparison group by the standard deviation \citep{harder2010propensity}. We see that balance improves substantially across all six confounders for all three implementation approaches (Figure~\ref{figure4}). 
\subsection{Sensitivity Analysis}
\begin{figure}[h]
\centering
\includegraphics[width=1\textwidth]{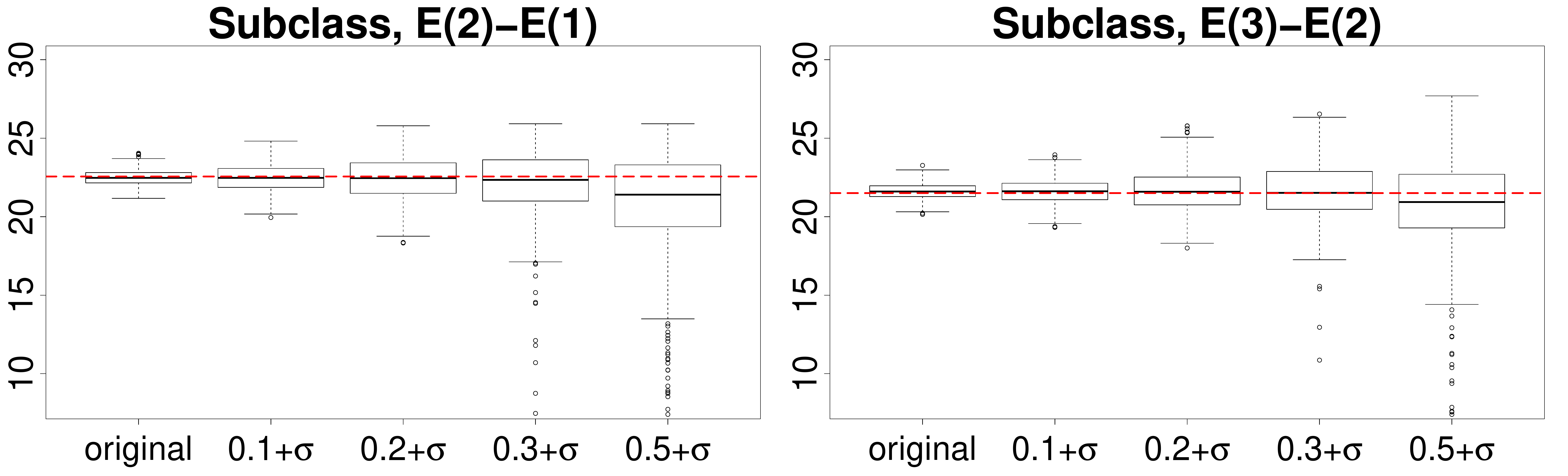}
\caption{Sensitivity analysis of the ATE estimates based on GPS approach using subclassification. Original represents the estimated ATE when the transportability assumption holds. Various violations of the transportability assumption are conducted, by sampling $\gamma$ in the RC model from a distribution with mean $\hat{\gamma}$ and augmented standard deviation. The red dashed line represents the true ATE. Results correspond to the default setting described in Table~\ref{tablep}. 
}
\label{figure6}
\end{figure}

We assess the sensitivity of the proposed approach to the transportability assumption by evaluating how well the approach performs under settings in which the true $\gamma$ is misspecified. We sample $\widehat{\gamma_{1,a}}$ from a normal distribution with mean $\hat{\gamma_1}$ and augmented standard deviation estimated by adding absolute values $0.1/0.2/0.3/0.5$ to the original standard deviation (0.0023) of $\hat{\gamma}_1$. $\hat{X}$ is then estimated by $\hat{X}= \hat{\gamma_0} + \widehat{\gamma_{1,a}}W + \mathbf{\hat{\gamma_{2}}}^T \mathbf{D}$. By adding this misspecification of the RC model, we artificially violate the transportability assumption and show how this violation could affect the ATE estimates. The results in Figure~\ref{figure6} show that for subclassification  even when the standard deviation is around 100 times higher than the original standard deviation, 
 the estimated ATE using GPS  is still robust. Not surprisingly, the variances of the estimated ATE increase under this extreme setting. The results from using GPS with IPTW and matching are also robust to the violation of transportability assumption (See Figure A2 in \ref{suppA}).

\subsection{Additional Simulations}
We conducted additional simulations (as described in Table~\ref{tablep}) with varying degrees of exposure error and confounding bias. Details on the additional simulations can be found in the \ref{suppA}. Briefly, we show that in a variety of settings our proposed RC-GPS approach can significantly reduce the bias of the estimated ATE. More specifically, when the correlation between true exposure $X$ and error-prone exposure $W$ is low the proposed approach significantly eliminates the bias if the RC model is correctly specified and has good fit. Yet the proposed approach is more sensitive to the correct specification of the RC model in these settings (See Figure A3 in \ref{suppA}).  When the RC model does not fit well the proposed method improves the bias in the ATE, but does not completely eliminate it (See Figure A4 in \ref{suppA}). When the true treatment effect is small the proposed approach  eliminates the bias if the RC model is correctly specified and has good fit (See Figure A5 in \ref{suppA}). When confounding is large the proposed approach  significantly eliminates the bias if the RC model is correctly specified and has good fit, yet the variance of the estimated ATE increases (See Figure A6 and A7 in \ref{suppA}). Lastly, when the RC model is non-linear with respect to error-prone exposure $W$ the proposed approach significantly eliminates the bias if the RC model is correctly specified and has good fit. Yet if we only use a linear model as our RC model (when the data generating mechanism was from a non-linear model), we do not see bias reductions using the proposed approach (See Figure A8 in \ref{suppA}). Overall, through the simulation study, we show the RC-GPS approach works remarkably well with varying degrees of exposure error and confounding bias.

\section{Data Application\label{datapp}}
We apply the proposed RC-GPS method to estimate the effect of long-term  \PM \ exposure on health outcomes. While \PM  \ concentrations are continuous, our interest is in comparing the effects of exposure in three categories based on pre-specified \PM \ cut-offs. 
The current long-term \PM\ standard
is annual mean of $12.0 \ {\rm \mu g/m^3}$, refer to National Ambient Air Quality Standards (NAAQS) Table \citep{NAAQS} 
There is a wide literature studying the effect of \PM \ exposures at these higher levels \citep{dockery1993association,beelen2014effects,kioumourtzoglou2016pm2}, 
yet limited literature is available on the effects of exposure at lower levels \citep{shi2016low,villeneuve2015long}.

Our interest is in estimating the exposure effects in the lower ranges, the results of which can help inform future policy regulations. Specifically, we consider two cut-offs; annual mean \PM \ levels of $8 $ and $ 10 \ {\rm \mu g/m^3}$, resulting in three exposure categories. Our main study population is Medicare participants across New England (VT, NH, CT, MA, RI and ME) from 2000 to 2012, and all-cause mortality is the outcome of interest. This study population includes a total of  3.3 million individuals with 24.5 million person-years of follow up, who reside in $2,202$ zip codes. 

\begin{figure}
\begin{minipage}{.49\columnwidth}
  \centering
  \includegraphics[width=1\linewidth,height=1\linewidth]{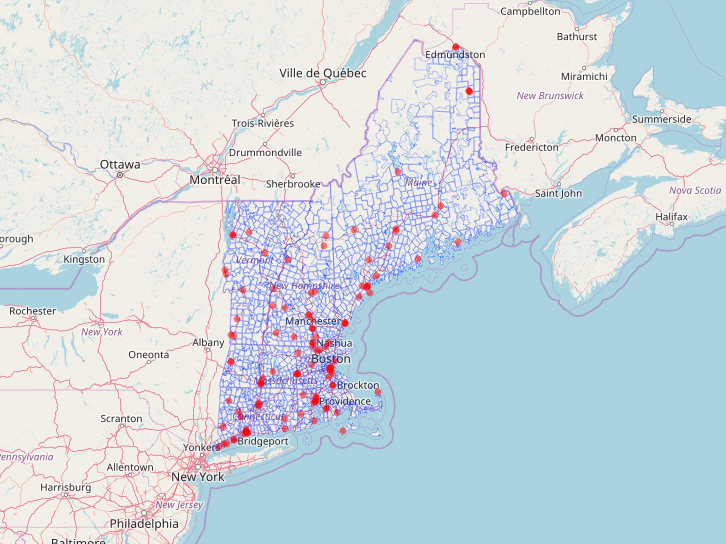}
  \captionof{figure}{Locations of monitor stations in New England (in red). Zip code areas are drawn in blue.}
  \label{figure8}
\end{minipage}\hfill
\begin{minipage}{.49\columnwidth}
  \centering
  \includegraphics[width=1\linewidth,height=1\linewidth]{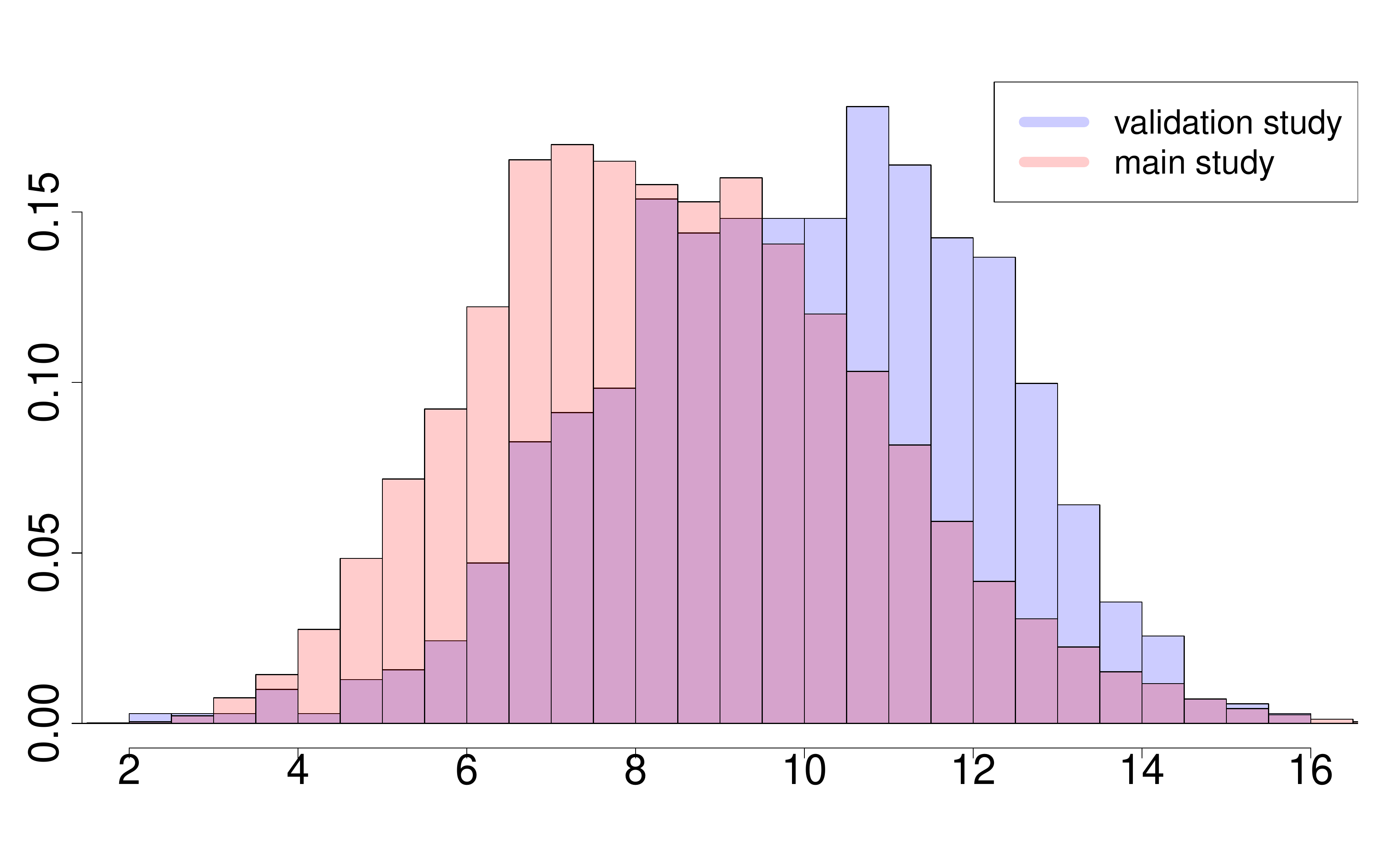}
  \captionof{figure}{The distribution of annual mean predicted \PM\ exposures in the main and the validation study across 13 years (2000-2012).}
  \label{figure9}
\end{minipage}
\end{figure}

\PM \ exposures are determined at each $1\textmd{km} \times 1\textmd{km}$ grid cell using a spatio-temporal prediction model which uses multiple different sources as input \citep{di2016assessing}. Although the prediction model performs well \citep{qd16}, 
there is still error associated with these predictions. 
For a subset of grid cells we have monitor stations that measure the actual observed \PM \ concentrations. We assume \PM\ concentrations monitored inside a grid cell are error-free exposures of the average concentrations in that grid cell, which is a reasonable assumption \citep{burton1996spatial,wilson1997fine,sarnat2010examination}. 
For New England we have $2,202$ zip codes covered by $217,660$ $1\textmd{km} \times 1\textmd{km}$ grids (main study). 
For a subset of these grids ($m=83$) within 75 zip codes,
we have actual \PM \ measured from monitoring stations (internal validation study). Figure~\ref{figure8} shows the locations of all 119 monitor stations in New England, 83 of which have actual \PM \ measures. 
Medicare data is available at the zip code level, yet \PM \ exposures are estimated at the grid level. To obtain annual average \PM \ at each zip code, we aggregate these gridded concentrations through area-weighted averages. The distributions of annual mean \PM\ exposures from the spatio-temporal prediction model in the main and the validation studies are compared in Figure~\ref{figure9}, showing that the monitors are not randomized across areas, i.e. they are more likely to be located at areas with higher \PM\ concentrations. 

\textit{RC Model.} The RC stage of our two-stage RC-GPS approach is implemented at the grid level. We have $217,660$ grids in the main study and $83$ grids in the internal validation study. After fitting the RC model,
we obtained estimates of the true \PM\ exposures at each grid cell in New England.
To improve the fit of the RC model we included 14 
meteorological variables as predictors, many of which were significant, with total cloud coverage ($p<0.001$) and total precipitation ($p=0.008$) as the most significant ones. The details of model fit are presented in the \ref{suppA}.
Subsequently, to obtain annual average \PM \ at the zip code, we aggregated the grid-level \PM \ exposures using area-weighted averages.
After the aggregation, we categorized the exposures into three categories, corresponding to exposure levels \PM\ $\leq 8\ {\rm \mu g/m^3}$, $8 <$ \PM\ $\leq 10\ {\rm \mu g/m^3}$, and \PM\ $> 10\ {\rm \mu g/m^3}$. 
For each calendar year that participants were at risk, their exposure was the annual average \PM\ for that year, based on their zip code of residence. 


\textit{GPS Model.} 
For the GPS model we include 16 area-level covariates as confounders. 
The GPS model is fitted using multinomial logistic regression with the 16 confounders. The details of the model fit are presented in the \ref{suppA}. 

\textit{Outcome Analysis.} Following the GPS implementations, we fit the outcome model using a stratified log-linear model with a person-time offset. We use 4 individual level covariates as stratification variables.
We do not include confounders in the outcome model and assume the GPS implementations fully adjust for all potential confounding. 
The incidence rate ratio (IRR) is estimated from the outcome model, and 95\% confidence intervals (CIs) were obtained by bootstrap with 100 replicates. We constructed the CIs using standard errors (SEs) estimated by bootstrap under the normality assumption, since estimation of a SE requires fewer bootstrapped replicates (25-200) than the direct estimation of the CI (1000-2000) \citep{efron1994introduction}. We conducted 100 replicates in the data application. 
\subsection{Data Analysis Results}
For each of the GPS implementations, we compare the estimated IRR using 1) a GPS approach using error-prone \PM \ exposures only, and 2) the proposed RC-GPS approach. 
To improve overlap, we trim the data to include only observations with GPS falling into the overlapping ranges of each GPS element among the different exposure subpopulations, which removes 1.7\% of the original data. 
For IPTW, we further set weights equal to 10 if the weights are greater than 10 \citep{harder2010propensity}, which truncates the weights of 3.1\% of the observations.
We conduct the outcome analysis based on the trimmed dataset. 
\begin{table}[H]
\centering
\caption{Data Application Results: ATE of long-term \PM\ exposure on mortality measured by incidence rate ratios (IRRs). Error-prone implements GPS approaches to adjust confounding based on error-prone exposures. RC-GPS is based on the proposed approach adjusting for measurement error by RC model and adjusting confounding using GPS approaches based on corrected exposures.  All 95\% confidence intervals were obtained by bootstrap.} 
\label{apptable}
\begin{tabular}{llll}
\hline
\noalign{\vskip 0.1cm}
\multicolumn{4}{ c }{Results for Exposure Levels \PM\ $\leq 8\ {\rm \mu g/m^3}$ vs. $8 <$ \PM\ $\leq 10\ {\rm \mu g/m^3}$} \\
     \hline
 & \multicolumn{3}{ c }{ATE [95$\%$ CI]} \\
  \hline
 &  \textbf{Subclassification}  &    \textbf{IPTW} &    \textbf{Matching}\\
\hline
GPS, Error-prone  & 1.013 [0.999, 1.029] & 1.031 [1.021, 1.042]& 1.020 [1.004, 1.036] \\
RC-GPS & 1.025 [1.006, 1.045]& 1.022 [1.007, 1.038]& 1.028 [1.012, 1.045]  
\\
 \hline
 \noalign{\vskip 0.1cm}
 \multicolumn{4}{ c }{Results for Exposure Levels \PM\ $\leq 8\ {\rm \mu g/m^3}$ vs. \PM\ $> 10\ {\rm \mu g/m^3}$} \\
     \hline
 & \multicolumn{3}{ c }{ATE [95$\%$ CI]} \\
  \hline
 &  \textbf{Subclassification}  &    \textbf{IPTW} &    \textbf{Matching}\\
\hline
GPS, Error-prone  & 1.015 [0.993, 1.037] & 1.050 [1.032, 1.068]& 1.018 [0.996, 1.040]
\\
RC-GPS & 1.035 [0.999, 1.072]& 1.030 [1.005, 1.056]& 1.035 [1.015, 1.055]   
\\
\hline
\end{tabular}
\end{table}
We see in Table~\ref{apptable} that the IRR estimates from the RC-GPS approach are consistent across all three implementations. The RC-GPS approach yields more pronounced point estimates compared to the error-prone implementation. 
The 95\% CIs are overlapping across all three implementations. For example, using matching we see a IRR of 1.028 under RC-GPS, for exposure category 1 vs. category 2, meaning a moderate exposure level of annual average \PM \ ($8 <$ \PM\ $ \leq 10\ {\rm \mu g/m^3}$) causes a 2.8\% increase in all-cause mortality compared to low exposure level of annual average \PM \ (\PM\ $ \leq 8\ {\rm \mu g/m^3}$). 
Using error-prone exposures we see less consistent results across the three different GPS implementations. The difference in results across the three GPS implementations indicates that these three approaches have different levels of sensitivity to measurement error. It is worth noting that trimming weights in IPTW will affect the estimate of ATE itself \citep{harder2010propensity}, thus possibly causes the differences in results between IPTW and the other two approaches.  

\begin{figure}[H]
\includegraphics[height=0.3\textheight,width=1\textwidth]{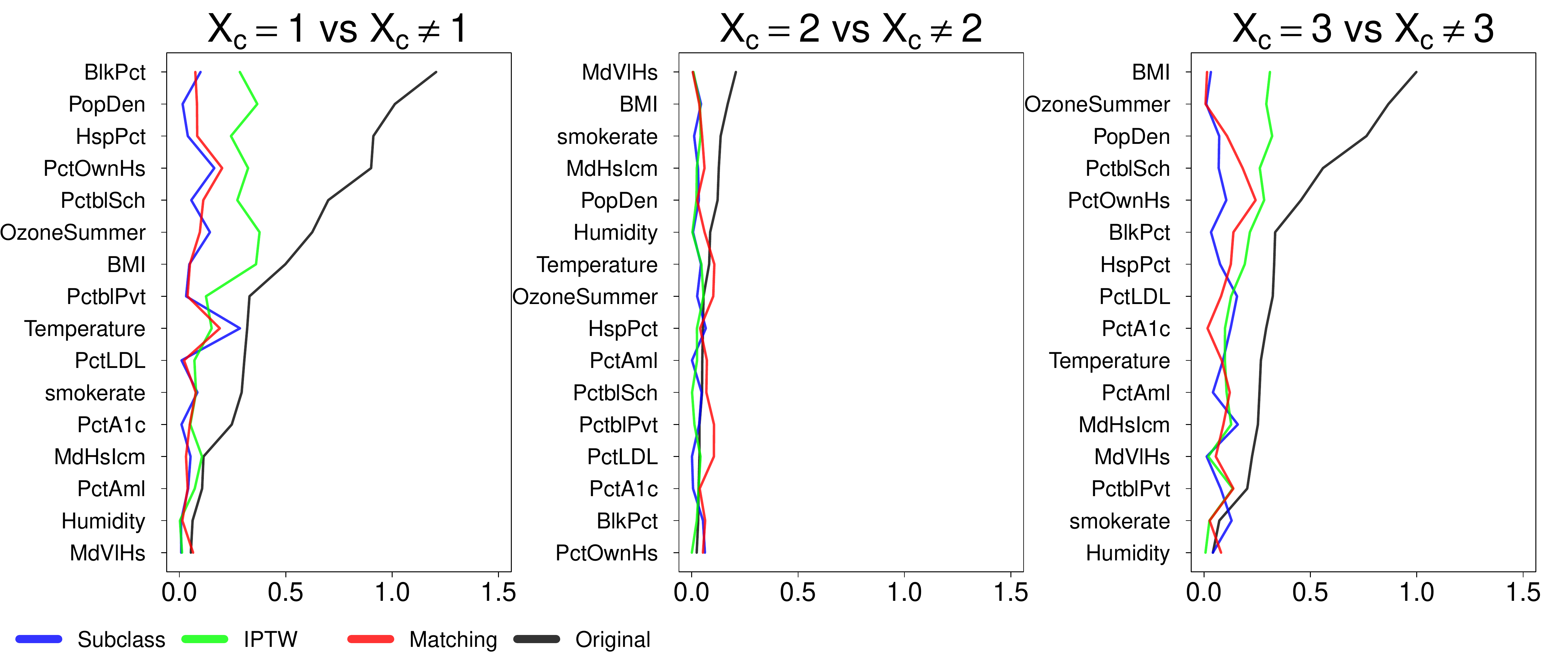}
\caption{Absolute Standardized Bias (ASB). Each panel represents the absolute standardized biases for each confounders, between subpopulation with $X_c=x$ and subpopulation with $X_c \neq x$ in original data (black) and after GPS implementations (colored). All three GPS implementations improve the covariates balances for most of confounders.}
\label{alance}
\end{figure}

We assess overlap by evaluating the distributions of the GPS elements for each exposure category as discussed in Section~\ref{simulations3}.
Comparing the figures before and after trimming (See \ref{suppA}), 
we notice the overlap assumption does not hold using the original data, but improves after trimming. This highlights the necessity of trimming in order to improve overlap. 
We assess the balance by calculating the ASB for each confounder for the different exposure categories (e.g. $X_c=1$ vs. $X_c\neq 1$) as discussed in Section~\ref{simulations3}. It is also important to note, that based on the calculations and figures the GPS implementations largely improve covariate balance for most of confounders (See Figure~\ref{alance} and Table A10 in \ref{suppA}). 

\section{Discussion\label{disc}}
We developed an innovative two-stage approach, RC-GPS, to estimate the average causal effect on a categorical scale in the setting of GPS analysis while correcting for measurement error in continuous exposures. Our simulation study shows that the proposed method has the potential to fully adjust for both the mismeasured exposure as well as confounding bias. We have also conducted sensitivity analyses and showed that the approach is robust under modest levels of model misspecification and assumption violations. 

The assumptions for the first component of our proposed approach, the RC model, are 1) transportability  and 2) non-differential measurement error (surrogacy). In our setting, we require the transportability of $E(X|W, \mathbf{D})$. Although this is not verifiable, it can be evaluated by sensitivity analysis, which we included as part of our simulation study in Section~\ref{simulations}. We see that in simulation scenarios considered, results are robust to the violation of this assumption. For the application, one should give careful thought about how likely this assumption will hold, and a sensitivity analysis to assess how the ATE varies for the violation of transportability assumption is recommended. 
The non-differential measurement error (surrogacy) assumption is believed to be held in many similar settings in 
air pollution studies \citep{dominici2000measurement,mak14_error,hart15_error}. 
We believe it holds, since in the context of air pollution applications, it is reasonable to assume that given the true exposures and confounders, the error-prone exposures do not provide any additional information on the health outcomes (i.e. mortality). In our study, we only consider the RC model as a linear regression model, though the form of RC model is not restricted to linear regression. However, since the RC model is only an approximation, fully adjusting for measurement error relies on assumptions, e.g. measurement error is ``small'' \citep{carroll1990approximate}, or outcome models without severe curvature \citep{carroll2006measurement}, if other forms of the RC model are used.

The regression of $X$ on $(W,\mathbf{D})$, is an art \citep{carroll2006measurement}, since true model of $X|W,\mathbf{D}$ can never be known. 
Therefore, assessing the robustness of the RC to model misspecification is very important. In summary, we considered three types of model misspecification for the RC model (\ref{eq1}) in simulations: 1) vary $\gamma$ in a correctly specified RC model, 2) exclude covariates $D$ associated with the measurement error in the RC model, 3) vary the true model structure of the RC model by introducing a non-linear (quadric) relationship between true exposure $X$ and error prone $W$. 
The first type of model misspecification introduces additional variability in estimating the coefficients $\gamma$, although it does not violate the true structure of the RC model (\ref{eq1}). Under this type of misspecification, we found that the RC-GPS maintains the capability to eliminate the bias, yet the variances of the estimated ATE increase. The second type actually violates the specification of the RC model (\ref{eq1}) since omission of the covariates $\mathbf{D}$ changes the mean function. 
We found that under this type of model misspecification, the RC-GPS approach still reduces the bias, although it does not completely eliminate it. The third type (shown in Figure A8 in \ref{suppA}) is a severe misspecification, since we introduce a non-linear relationship between $X$ and $W$, which we ignore when we fit the RC model. 
We present this type of extreme model misspecification to show that under such a severe case, bias reduction is not guaranteed. Correct specification of the RC model, however, would still eliminate the bias. It is highly recommended, therefore, that goodness of fit is assessed and sensitivity analyses performed to best characterize the functional form of the RC model in real-life applications.



For the GPS implementation, there are three main assumptions 1) no-interference, 2) overlap, and 3) weak unconfoundeness. The no-interference assumption is a fundamental assumption in the potential outcome framework. In the air pollution context it could be violated as exposure in the current period could affect mortality in subsequent periods \citep{baccini2017assessing}. \cite{baccini2017assessing} argue that by enlarging the time window of exposure averages the no-interference assumption is more likely to hold. In the data application we consider long-term annual mean exposures (rather than short-term daily exposures), which will likely increase the validity of the assumption. 

As we saw in the data application, there are settings in which the overlap assumption does not hold. 
A common approach, and also the approach we used in data application, to improve overlap is trimming the sample  by disregarding subjects with low and high values of GPS elements \citep{crump2009dealing}. Yet the limitation is that by doing so we alter the target population, and thus the target estimand. The estimated ATE based on the trimmed sample, can deviate from the quantity of interest, i.e. the ATE for the whole population, thus we need to carefully think about the generalizability of our results. In our data application, 1.7\% of the observations were trimmed. We compare population characteristics in the entire population and trimmed population, and there is no evidence that those two populations are significantly different (See Table A8 in the \ref{suppA}). 

The unconfoundeness assumption is not verifiable, since data is always uninformative about the distribution of counter-factual outcome for unreceived exposures, yet this is a common assumption in propensity score analysis. In our setting, we only assume weak unconfoundeness, which only requires the potential outcome for each category of exposure and the exposure to be assigned at the corresponding category, are pairwise independent conditional on all potential confounders \citep{imbens2000role}. 
It is weaker than strong unconfoundeness defined in \citet{rosenbaum1983central}, which requires the joint distribution of potential outcomes to be independent with the assignment mechanism for all exposures conditional on all potential confounders. The limitation of weak uncoundoudedness is that we are not able to estimate ATE for subpopulations, i.e. the ATE in populations with exposure category 1 and 2 only. 
However, the interest is usually in estimating the ATE for the whole population, which can be estimated under weak unconfoundeness.

For inference, the main strategy is to use bootstrapping to obtain the CIs, and the validity of inference is guaranteed by the validity of the bootstrap procedures. The RC stage in most scenarios does not introduce a substantial amount of additional variability in effect estimates, except when both the RC model lacks fit and the validation size is small (See Table A2-A3 in \ref{suppA}). In the data application, the results show (slightly) wider CIs for RC-GPS compared to the naive GPS estimates which do not consider the RC correction (Table~\ref{apptable}). The reason that we obtain wider CIs for RC-GPS compared to GPS without correction is that: 1) the validation size is relatively small, 2) the propagation of the uncertainty in the RC stage, 3) in our application, Medicare data is available at the zip code level, but \PM\ exposures are estimated at the grid level. 
To obtain annual average \PM\ at each zip code, we aggregate these gridded concentrations through area-weighted averages. The aggregation procedure itself could potentially introduce a lot variability as well. Explicitly, the aggregation procedure may amplify the uncertainty during the estimation of RC models. The GPS estimation stage, in general, does not increase the variability in effect estimates compared to those using the true GPS, since \citet{lunceford2004stratification,abadie2016matching} proved that for all three types of PS implementations using the estimated PS is more efficient than using the true PS in large samples, and the arguments in \cite{imai2004causal} confirmed the conclusion can be extended to GPS settings. We found similar finite sample properties in simulations (See Table A4-A5 in \ref{suppA}). 

This data application illustrates the ability of the proposed RC-GPS approach to estimate corrected causal effects between long-term \PM \ on a categorical scale and all cause mortality. There are a few potential limitations in our analysis of the data application. The first is that the validation data is not a random sample of main study, since monitor locations are not randomized across areas, i.e. more monitors are within urban areas, which could impact the transportability assumption. However, by conditioning on additional covariates in the RC model, such as weather variables which explain geographical heterogeneity between grid cells, the transportability assumption is more likely to hold. The second is that the RC model is not guaranteed to be correctly specified or have a good fit, due to both lack of potential predictive covariates and modeling assumptions. 
The third is that the estimated GPS can be biased since it relies on correct specification of the GPS model, and is not robust to unmeasured confounding. The correctly specified GPS maintains a balancing property as described in Section~\ref{simulations3}. In the data application, we assessed balance and there was some evidence of imbalance. Even with some evidence of imbalance, the advantage of GPS approaches is that they are more robust to outcome model misspecification compared to fitting outcome models with confounders as covariates. 
The fourth is that overlap may still be limited even after we trim the data. 
There is always a trade-off between guaranteeing overlap and trimming out samples excessively, thus modifying the population. The fifth is the cut-off selection. In the data application, 
cut-offs were selected from a policy perspective. The first \PM\ cut-off was selected at $10\ {\rm \mu g/m^3}$, which is the current Air Quality Guideline proposed by the World Health Organization (WHO) for annual \PM\ concentrations  \citep{WHO}. Currently, in the US the NAAQS is $12\ {\rm \mu g/m^3}$ \citep{NAAQS}. 
Understanding, therefore, the effect of \PM\ exposure on mortality in the US population at lower levels, like the WHO
guideline or even lower, is of great interest, as it can inform regulatory action. This is also the motivation behind including an additional cut-off at $8\ {\rm \mu g/m^3}$ annual mean. From a statistical perspective, potential cut-off choices could be driven by modeling assumptions. Specifically, cut-offs can be selected to ensure that the overlap assumption holds for valid causal inference. Based on the distribution of annual mean \PM\ exposures, the chosen $8\ {\rm \mu g/m^3}$ and $10\ {\rm \mu g/m^3}$ cut-offs divide the units into three categories approximately evenly, which arguably is more likely to ensure overlap. We have also conducted some sensitivity analysis to evaluate the overlap assumption (See Table A9 in the \ref{suppA}); the current cut-off points ($8\ {\rm \mu g/m^3}$ and $10\ {\rm \mu g/m^3}$) provided the best overlap. 

Although air pollution has motivated our application, the proposed RC-GPS  approach is not limited to one specific area. One other potential application of this approach could be in the setting of clinical studies of biomarkers. Commonly, in clinical trials, we are interested in the dose response of categorical levels of drug doses, e.g. Vitamin D supplements, yet in order to evaluate this treatment in human subjects, we measure the biomarkers, e.g. blood levels of Vitamin D. In this setting we might have an error-prone continuous treatment for each patient based on blood work from routine medical examinations, and an internal subset of samples for which we know the true treatment based on blood work from a more robust central laboratory. The accurate measures of blood samples through a standard central laboratory are costly and infeasible for every patient, and therefore we obtain this gold-standard measurement only for a subset of patients (internal validation study).
A detailed example can be found in \citet{gail2016calibration}. In such study designs, one can fit a RC model to estimate true treatments for each patient. After that, one can use GPS based on the estimated true treatments in the main study, in order to obtain causal effects for various vitamin D doses, to determine the most effective dose of Vitamin D supplements.

The RC-GPS approach introduced in this paper is the first approach which allows for the correction of exposure error in both design and analysis phases using GPS, and assesses covariate balance through standardized bias in the context of GPS for categorical exposures. It can be further generalized to estimate causal effects on a continuous scale rather than categorical to answer different scientific questions. Simulations have shown the proposed approach is robust and we are optimistic about the adaption of the approach to various research areas.






%


\section*{Acknowledgements}
This work was made possible by the support from the Health Effects Institute (HEI) grant 4953-RFA14-3/16-4, National Institute of Health (NIH) grants R01 GM111339, R01 ES024332, R01 ES026217, R01 ES028033, R01 MD012769, DP2 MD012722, NIH/National Institute on Minority Health and Health Disparities (NIMHD) grant P50 MD010428, and USEPA grants RD-83587201-0, RD-83615601. The contents are solely the responsibility of the grantee and do not necessarily represent the official views of the funding agencies. Further, funding agencies do not endorse the purchase of any commercial products or services related to this publication. Research described in this article was conducted under contract to the HEI, an organization jointly funded by the USEPA (Assistance Award No. R-83467701) and certain motor vehicle and engine manufacturers.
The contents of this article do not necessarily reflect the views of HEI, or its sponsors, nor do they necessarily reflect the views and policies of the EPA or motor vehicle and engine manufacturers. The computations in this paper were run on the Research Computing Environment (RCE) supported by the Institute for Quantitative Social Science in the Faculty of Arts and Sciences at Harvard University. We thank the referees for careful readings and thoughtful comments.

\begin{supplement}
\sname{Supplement}\label{suppA}
\stitle{Supplementary Materials}
\slink[url]{http://www.e-publications.org/ims/support/dowload/imsart-ims.zip}
\sdescription{Supplementary Figures, Tables and Text. \\ R code for simulations is available at \href{https://github.com/wxwx1993/RC-GPS}%
{https://github.com/wxwx1993/RC-GPS}.}
\end{supplement}

\bibliographystyle{imsart-nameyear}
\bibliography{xiao_wu}

\end{document}